# Dispersion of critical rotational speeds of gearbox: effect of bearings stiffnesses

F. Mayeux, E. Rigaud, J. Perret-Liaudet
Ecole Centrale de Lyon
Laboratoire de Tribologie et Dynamique des Systèmes
Batiment H10, 36, avenue Guy de Collongue
69134 ECULLY Cedex, France
e-mail: francois.mayeux@ec-lyon.fr

## Abstract
Noise measurement on a population of a gearbox manufactured in large number reveals a high variability principally due to tolerances on each design parameter of the gear. Gearbox noise results mainly from vibration of its housing excited by the transmission error. High dynamic mesh load, high vibratory response of housing and high acoustic level of the gearbox can be observed for some critical rotational speeds. These ones correspond to excitation in a resonant manner of some particular modes which store a high potential energy at the meshing stiffness. Variability of the associated critical modes depend on tolerances affected to gear design parameters. It also depends on bearings stiffnesses dispersion. The physical origins of the bearings stiffnesses variability are preloads. Bearings stiffnesses and tolerances affect the characteristics of the meshing stiffness and the dynamic transfer from dynamic meshing force to the housing vibratory response. The purpose of this paper is to study effect of the bearings stiffnesses variability by estimating the first two moments of statistical distribution of critical modes.

## 1   Introduction

The vibratory and acoustical behaviour of gearboxes results from numerous sources [1]. Among these, it is generally admitted that the main source is the static transmission error under load (STE) [1,2]. STE is mainly governed by periodic components at the meshing frequency due to (1) elastic deflections of gear teeth under load (periodic meshing stiffness), (2) teeth geometry modifications, (3) manufacturing errors and (4) shaft misalignments. Under operating conditions, STE generates dynamic mesh forces leading to dynamic forces and moments transmitted through bearings, housing vibration and noise.

Further, critical rotation speeds associated with high dynamic mesh forces and high noise levels, correspond to the excitation of some critical modes having a high potential energy stored by the meshing stiffness [3]. These critical speeds are mainly controlled by the time-average meshing stiffness and bearings stiffnesses.

At last, considering gearbox manufactured in large number, we observe dispersion of critical speeds and excitation levels mainly due to the variability of STE, meshing stiffness and bearings stiffnesses. Sources of dispersion result mainly from geometry errors authorised by designers who introduce necessary tolerances.

In this context, the aim of this paper is to deal with some results about variabilities of peak to peak STE, time-average meshing stiffness, bearings stiffnesses and critical speeds. Statistics are obtained from a modified Taguchi's method. Introduced sources of variability are the helix errors, the shaft misalignments and the bearing's preload on each shaft. One can remark that bearing's preloads play a game not only on the natural modes of the overall model (effect of stiffness matrix) but also on excitation source (effect of misalignment between mating wheel induced by static bearing deformations).

## 2   Computational methods

### 2.1   Modified Taguchi's method

A modified Taguchi's method allows to estimate in a very simple way the statistical moments of a function of multiple random variables whose probability densities are known [4]. Statistical moments are estimated from numerical integration of Gauss quadrature type. Then, the response function is calculated for a relatively short number of samples, judiciously chosen. For each uncertain variable, a number of samples up to three is necessary to take into account the eventual non-linear behaviour of the response function. Precision increases rapidly with this number of samples. More precisely, the number of samples is equal to the product of the number of levels chosen for each factor. In this study, we used 3 levels per factor, so 27 samples are needed to treat the three retained random parameters. Finally, the principal advantages of this method are the ease of its numerical implementation and its short computing time. Efficiency of the method is clearly demonstrated in the case of overall gearbox dynamic and noise prediction [5]. Unfortunately, the modified Taguchi's method can't provide probability density function. Furthermore, from the computed samples, one can also estimate the influence of the uncertain parameters by using standard variance analysis.

## 2.2 Calculation of the STE

The STE is obtained by solving the static equilibrium of the gear pair for a set of successive rotational positions of the driving wheel [6]. For this end, the theoretical tooth contact lines contained in the action plane are discretized in a some number of slices. At each slice, the unknown contact load is assumed to be positive or zero (column vector $\mathbf{P}$). The calculation requires knowledge of the compliance square matrix, $\mathbf{C}$, acting between slices. This matrix can be obtained from a previous 3D finite element model of the mating teeth. It requires also the knowledge of manufacturing errors which are introduced as a column vector $\mathbf{e}$ of initial gap at each slice. Finally, the misalignment measured in the plane of pressure, $(\phi_1-\phi_2) = \phi$, which is induced by the shafts, bearings and housing deformations, is taken into account separately from vector $\mathbf{e}$. The static teeth contact problem to solve can be written as follows:

$$\mathbf{C}\,\mathbf{P} = d.\mathbf{i} + \mathbf{e} + \phi.\mathbf{g} \quad \text{and} \quad {}^t\mathbf{P}.\mathbf{I} = N$$
$$\text{with} \quad P_j \geq 0 \qquad (1)$$

Here, d is the unknown STE, $\mathbf{i}$ is a column vector of ones, $\mathbf{g}$ is a column vector which localizes slices in the action plane and N is the total normal force transmitted through the action plane. Under some rearrangements, equation (1) is solved by using a modified simplex method [7]. The computation directly allows to obtain the STE and the load distribution along contact lines. Then, the generalized forces $\mathbf{F}_k$ acting at each center $O_k$ of the mating wheels can be obtained from the load distribution. Finally, the meshing stiffness can be computed considering an increment of the normal force (numerical derivative around the static condition). As the STE, meshing stiffness is a periodic function in relation with the cyclic variation of the number of in contact teeth pair.

## 2.3 Calculation of bearings stiffnesses

Consider the static equilibrium of an elastic shaft supported by two rolling element bearings A and B which are mounted in a rigid housing. Assume the shaft subjected at a point G to an arbitrary generalized force vector $\mathbf{F}$ which induces generalized reaction force vector at each bearing, $\mathbf{R}_A$ and $\mathbf{R}_B$. By introducing the shaft as two linear FEM super-elements with their stiffness matrices assumed to be known, $\mathbf{K}$, one can obtain the shaft static equilibrium as follows:

$$\begin{bmatrix} \mathbf{K}_{AA} & \mathbf{K}_{AG} & 0 \\ \mathbf{K}_{AG} & \mathbf{K}_{GG} & \mathbf{K}_{BG} \\ 0 & \mathbf{K}_{BG} & \mathbf{K}_{BB} \end{bmatrix} \begin{Bmatrix} \mathbf{x}_A \\ \mathbf{x}_G \\ \mathbf{x}_B \end{Bmatrix} - \begin{Bmatrix} \mathbf{R}_A \\ 0 \\ \mathbf{R}_B \end{Bmatrix} = \begin{Bmatrix} 0 \\ \mathbf{F} \\ 0 \end{Bmatrix} \qquad (2)$$

$\mathbf{x}_A$, $\mathbf{x}_B$ and $\mathbf{x}_G$ are respectively the generalized displacement vectors at each point A, B and G. Now, consider the equilibrium of the rolling element bearing A. One can obtain equations which relate the generalized inner ring displacement $\mathbf{x}_A$ to the bearing reaction force $\mathbf{R}_A$ [8] by considering the non linear forces acting on each rolling element, i.e. $\mathbf{Q}_j = \mathbf{Q}_j(\mathbf{T}_j\,\mathbf{x}_A)$. The non linearity results from the elastic contact law (rings are assumed to be rigid). $\mathbf{T}_j$ is a simple transformation matrix allowing to precise local displacement of the inner ring relative to the outer ring at the rolling element. Bearing reaction force is then given by:

$$\mathbf{R}_A = \sum_{j=1}^{n} \mathbf{T}_j \mathbf{Q}_j(\mathbf{T}_j \mathbf{x}_A) = \mathbf{R}_A(\mathbf{x}_A) \qquad (3)$$

Reaction force at bearing B is obtained in a similar manner. Substituting $\mathbf{R}_A$, $\mathbf{R}_B$ in equation (2) leads to:

$$[\mathbf{K}]\{\mathbf{x}\} - \{\mathbf{R}(\{\mathbf{x}\})\} - \{\mathbf{F}\} = \{0\} \qquad (4)$$

This non linear equation is solved by using a Newton-Raphson method. Further, if necessary, elasticity of the housing can be introduced in the same manner as shaft elasticity. The calculation directly allows to obtain the bearing reaction forces and the bearing stiffness matrices. It allows also to obtained the generalized displacement of each toothed wheel and then the in plane action misalignment φ.

## 2.4 Critical modes

The computation of the vibratory response of the gearbox induced by the STE requires a finite element modeling of all its components, i.e. gear, shafts, bearings and housing. A specific 12x12 stiffness matrix couples the axial, radial and rotational motions of the driven wheel and those of the driving one. This matrix is derived from the above calculation and from the geometrical characteristics of the gear pair [3,9]. Further, specific 5x5 stiffness matrices are introduced for modeling the bearings. Considering the average time meshing stiffness, one can obtained the eigenmodes of the overall gearbox. In order to extract the critical modes, an energetic coefficient, $\rho_i$, representing the elastic storage in the meshing stiffness for each mode is calculated. The higher $\rho_i$ is, the more critical the mode is [3,9].

## 2.5 Critical speeds and associated ranges

Critical speeds correspond to the excitation by STE of the above critical modes such that their eigenfrequencies are equal to the meshing frequency or its harmonics:

$$n Z_j f_j = f_i \quad (5)$$

where n is an integer, $Z_j$ is the number of teeth of the wheel number j, $f_j$ is its rotation frequency and $f_i$ is the frequency of the critical mode number i. For a critical speed expressed in rpm:

$$N_j = 60 f_i / n Z_j \quad (6)$$

Finally critical speed ranges are obtained with Tchebycheff's inequality.

## 2.6 Computation of the vibratory response

By linearizing equations around the static equilibrium, the vibratory response of the modeled geared system is governed by the following system of linear differential equations with periodic coefficient:

$$\mathbf{M}\ddot{\mathbf{X}} + \mathbf{C}\dot{\mathbf{X}} + \mathbf{K}\mathbf{X} + k(t)\mathbf{D}\mathbf{X} = \mathbf{E} \quad (7)$$

In this equation, **X** is the vibratory response of the gearbox, **M** and **K** are the mass and stiffness matrices provided by the finite element method (including the bearings stiffness matrices), k(t) is the periodic meshing stiffness, **D** is a matrix which couples the two toothed wheels and **E** is the generalized force vector which results from the STE. Matrix **C** represents damping which is introduced later into every modal equation. The vibratory response is compute in the average modal base by using an efficient method described in [10] and named "Iterative Spectral Method". It directly gives complex spectrum of the response at each degree of freedom of the modeled gearbox transmission. The computation time associated with this method is about 100 times shorter than that associated with standard numerical time integration schemes. Finally, the dynamic mesh load is deduced from the vibratory response.

## 3 The studied gearbox

### 3.1 Description of the gearbox

A single-stage involute helical gear has been investigated. Description of the main characteristics is given in Table 1. The design load for this gear pair is of 4000 N, which corresponds to a torque of 152 Nm. In this analysis one of fourth of the design load is considered. A standard profile correction is introduced consisting on tip relieves of 6 μm. The gearbox is fitted out with four ball bearings which have a 20° nominal contact angles. Shafts are 40 mm diameter, and 60 mm length. Gears are localized in the middle shaft. The housing has a shape of a rectangular prism (190x120x100 mm, see figure 1). The finite element model of the housing is shown in figure 1. The overall gearbox model has 1600 elements and 10000 degrees of freedom. 600 master degrees of freedom are retained. The gearbox has 60 eigenmodes in the frequency range 0-5000 Hz.

|  | Pinion | Driven wheel |
|---|---|---|
| Number of teeth | 28 | 58 |
| Base radius (mm) | 35.8 | 74.2 |
| Normal module (mm) | 2.5 | |
| Pressure angle | 20° | |
| Helix angle | 25° | |
| Face width (mm) | 20 | |
| Centre distance (mm) | 120 | |

Table 1: Main geometrical characteristics of gear

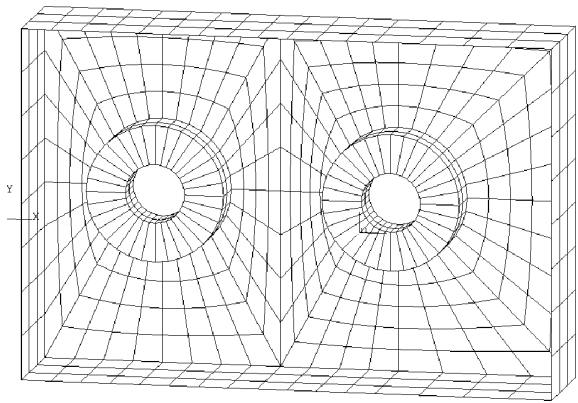

Figure 1: Gearbox's housing (190x120x100 mm)

## 3.2 Tolerances

In this study, lead error, shaft misalignments (deviation and inclination) are assumed to be random variables in the tolerance ranges. The tolerance are chosen considering the quality class 7 of the AFNOR French Standard NF E23-006. This quality class is often used in industrial applications (gearbox, machine tool,…). Preloads on each shaft are also assumed to be random variables. All the parameters are assumed to be truncated gaussian parameters with a tolerance equal to 6 times the standard deviation. The two first statistical moments for these parameters are given in Table 2. Mean values are considered as the deterministic case.

|  | Mean value | Standard deviation | Tolerance range |
|---|---|---|---|
| Lead error | 0 µm | 4 µm | ±12 µm |
| Deviation | 0 µm | 2 µm | ±6 µm |
| Inclination | 0 µm | 4 µm | ±12 µm |
| Preload 1 | 8 µm | 2.67 µm | ±8 µm |
| Preload 2 | 8 µm | 2.67 µm | ±8 µm |

Table 2 : The 2 first statistical moments and tolerances of the random parameters

Helix error is calculated from lead error, deviation and inclination. Assuming that these parameters are statistically independent, the induced helix error is also gaussian with a zero mean value and a standard deviation equal to 4.7 µm. Finally, the three studied independent factors are the overall helix error H, and the two preloads P1 and P2.

## 4   results

### 4.1   STE, meshing stiffness and bearing stiffness variabilities

In the figure 2, STE is plotted over two meshing periods for two set of parameters H, P1, P2 inside the tolerance ranges.

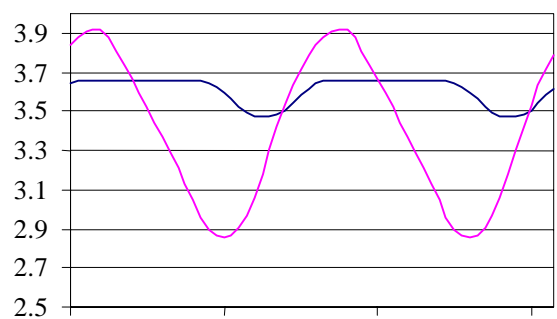

Figure 2: STE (µm) over two meshing periods

As you could see STE could be strongly modified for different sets of parameters inside the tolerance ranges. From knowledge of manufacturing tolerances, Taguchi's method allows to calculate mean value and standard deviation of the peak to peak STE, fundamental and first STE harmonics. Results are given in table 3.

| STE | Peak to peak | $f_e$ | $2 \times f_e$ |
|---|---|---|---|
| Mean value | 10.2 µrad | 4.6 µrad | 0.96 µrad |
| Standard deviation | 5.4 µrad | 4.0 µrad | 0.04 µrad |

Table 3 : Statistical moments of the peak to peak STE, fundamental and first STE harmonics

One should have high variability on noise level response because the STE constitutes the main excitation source.
Statistical moments of the meshing stiffness are given in table 4. Also deterministic case obtained from mean value of parameters H, P1 and P2 is given in this table.
By using Tchebycheff's inequality, meshing stiffness has a minimum probability equal to 96% to be in the range from 255 N/µm to 350 N/µm. Although this range is overestimated, its length remains unnegligible.

|  | Meshing stiffness |
|---|---|
| Deterministic | 309.3 N/µm |
| Mean value | 302.9 N/µm |
| Standard deviation | 9.5 N/µm |

Table 4 : Meshing stiffness

Concerning bearings stiffness variabilities, we chose to show only three diagonal terms (i.e. $K_{xx}$, $K_{yy}$ and $K_{zz}$) of the stiffness matrix for one of the four bearings. Results are given in table 5. As we can see, variabilities is not very pronounced.

|  | $K_{xx}$ | $K_{yy}$ | $K_{zz}$ |
|---|---|---|---|
| Deterministic | 48.7 N/µm | 82.7 N/µm | 14.8 N/µm |
| Mean value | 49.2 N/µm | 82.5 N/µm | 15.4 N/µm |
| Standard deviation | 0.8 N/µm | 0.2 N/µm | 0.9 N/µm |

Table 5 : Bearing stiffnesses

## 4.2 Critical speeds variability

From the modal analysis, we have found three main critical modes. Deterministic values, mean values and standard deviations for their natural frequencies $f_{Ci}$ and their energetic coefficient $\rho_i$ are given in table 6.

|  | Deterministic | Mean value | Standard deviation |
|---|---|---|---|
| $f_{C1}$ | 3994 Hz | 3992 Hz | 3.2 Hz |
| $\rho_1$ | 5.3 % | 4.5 % | 0.9 % |
| $f_{C2}$ | 4416 Hz | 4403 Hz | 25 Hz |
| $\rho_2$ | 9 % | 23 % | 20 % |
| $f_{C3}$ | 4486 Hz | 4470 Hz | 24 Hz |
| $\rho_3$ | 54 % | 40 % | 20 % |

Table 6 : Critical eigenfrequencies and their energetic coefficients

As we can see, the variability of the first critical mode is weak. In contrary, variabilities of the second and the third critical modes are strong. Considering standard deviations of natural frequencies and Tchebycheff's inequality leads to a range of 250 Hz for a 96% probability to be in this range. Furthermore, one can observe a very large standard deviation of $\rho_2$ and $\rho_3$. In fact, there exists an energetic transfer between the two modes with a total energetic coefficient ($\rho_2 + \rho_3$) remaining constant to 63%. This leads to equal standard deviations for $\rho_2$ and $\rho_3$. So critical speeds take place in a large range of rotational speeds. Considering excitation of these modes by the first harmonic of the STE and using the Tchebycheff's inequality with a 96% probability leads to a meshing frequency over 312 Hz (see figure 3). High variability of this critical speed and transfer between modes are illustrated in figure 4 where we show the dynamic mesh load for three sets of parameters H, P1 and P2.

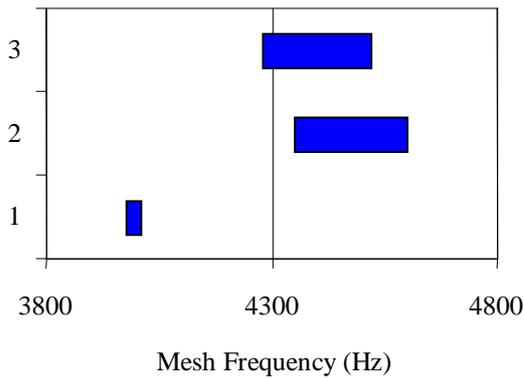

figure 3 : range of critical speed

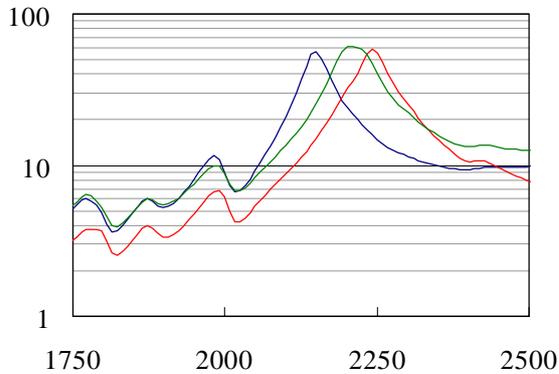

figure 4 : Dynamic mesh force (N) vs mesh frequency (Hz)

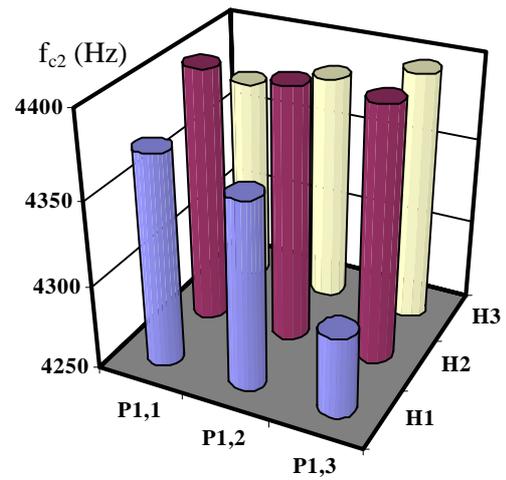

figure 5 : Matrix representation of the factorial experiments H-P2 with P1 = 3.4 μm.

### 4.3 Influence analysis

It is interesting now to estimate influence of each fault on the dispersion. As an example, figure 5 shows in a graphical manner the nine samples which are obtained by fixing the preload P1. More precisely, every point corresponds to the value of the frequency of the second critical mode for one attempt of the factorial experiments. Evolutions of the natural frequency are quite different when we change the level of the helix error. This result demonstrates clearly the importance of interaction between helix error and preload. Further, the helix error is a dominant parameter, while preload can also play a relatively significant role. Also, one can notice that, when the helix error is zero, the natural frequency has very little variations. Finally, these results have been observed for the other output parameters.

## 5 Conclusion

In this study, we used statistical methods to describe the variability of the critical speeds of a gearbox fitted out with a single helical gear pair.

Among all, the geometry faults of a geared system being able to contribute to the variability, the analysis focused on three parameters : (i) total helix error that includes misalignments of the shaft and the lead error, (ii) preload on the first shaft, and (iii) preload on the second shaft.

Obtained results lead to the following main conclusions:
- Large variability of the static transmission error and meshing stiffness has been obtained.
- In contrary, bearings stiffnesses present in the studied case little dispersion.
- Large variability of critical speeds (speed range and levels) has been obtained. This results from variability of critical modes with shape modifications and energy transfer between one mode to an another mode.

In future, it seems of interest to analyze effect of the design (bearing types, shaft length, housing, gear types,…) on the influence of the preload.

## Acknowledgements

The author would like to thank the Region Rhône-Alpes community and the SNR Roulements Company for supporting this work.

# References

[1] D. Welbourn, Fundamental knowledge of gear noise – a survey, Conference on Noise and Vibrations of Engines and Transmissions, Cranfield Institue of Technology C177/79, 9-29, 1979.

[2] D. Rémond, P. Velex, J. Sabot, Comportement dynamique et acoustique des transmissions par engrenages. Synthèse bibliographique. Publication du CETIM, 1993 (in French).

[3] E. Rigaud, J. Sabot, J. Perret-Liaudet, Approche globale pour l'analyse de la réponse vibratoire d'une transmission par engrenages. Revue Européenne des Elements Finis, 9, 315-330, 2000 (in French).

[4] J.R. D'Errico and J.R. Zaino, Statistical tolerancing using a modification of Taguchi's method, Technometrics 30(4), 397-405 1988.

[5] N. Driot, E. Rigaud, J. Sabot and J. Perret-Liaudet, Allocation of Gear Tolerances to Minimize Gearbox Noise Variability, Acta Acustica, 87, 67-76, 2001.

[6] E. Rigaud, D. Barday, Modelling and Analysis of Static Transmission Error. Effect of Wheel Body Deformation and Interactions between Adjacent Loaded Teeth. 4th World Congress on Gearing and Power Transmission, Paris, Vol. 3, 1961-1972, 1999

[7] T. Conry and A. Seireg, A Mathematical Programming Method for Design of Elastic Bodies in Contact, Journal of Applied Mechanics 93(1), 387-392, 1971.

[8] J. M. De Mul, J. M. Vree, D. A. Maas, Equilibrium and associated load distribution in ball and roller bearings loaded in five degrees of freedom while neglecting friction, Journal of Tribology 111, 140-155, 1989.

[9] E. Rigaud, J. Sabot and J. Perret-Liaudet, Effect of Gearbox Design Parameters on the Vibratory Response of its Housing, 4th World Congress on Gearing and Power Transmission, Paris, 3, 2143-2148, 1999.

[10] J. Perret-Liaudet, An original Method for Computing the Response of a Parametrically Excited Forced System, Journal of Sound and Vibration, 196(2), 165-177, 1996.